# The effect of Zn substitution on the state of oxygen deficiency and hole concentration in $Y_{1-x}Ca_xBa_2(Cu_{1-y}Zn_y)_3O_{7-\delta}$

S. H. Naqib[*a,b]

[a]*Department of Physics, Rajshahi University, Rajshahi 6205, Bangladesh*
[b]*IRC in Superconductivity, University of Cambridge, Madingley Road, Cambridge CB3 0HE, UK*

**Abstract**

We have investigated the effect of Zn substitution on the hole content, p, and the oxygen deficiency, δ, for a series of high-quality crystalline c-axis oriented thin films and polycrystalline $Y_{1-x}Ca_xBa_2(Cu_{1-y}Zn_y)_3O_{7-\delta}$ compounds over a wide range of composition (x, y, and δ). p was determined from the room-temperature thermopower, *S[290K]*. When annealed at a given temperature and oxygen partial pressure, a small but systematic variation in *S[290K]* was observed with Zn, for all the samples. From the analysis of the temperature derivative of resistivity, dρ/dT, at 300K, which varies almost linearly with p, irrespective of the composition of the sample, and the thermo-gravimetric analysis (TGA), we have found that annealing under identical conditions makes the Zn-substituted compounds somewhat more oxygen deficient compared to the Zn-free ones.

## 1. Introduction

Impurity substitution has proven to be a powerful probe of the bulk behavior of complex many-body systems, and for cuprate superconductors it is no exception. $T_c$ suppression, the increase of the residual resistivity, the temperature dependence of the penetration depth, the impurity-induced increase of electronic density of states in the low temperature, and impurity-related local bound states seen in STM are just some of the phenomena that have been investigated through impurity studies, and a wealth of information has been garnered [1-4]. Zn is among the most widely substituted impurity for Cu in cuprate superconductors. There are several advantages of using Zn, namely: (i) It mainly substitutes the in-plane Cu(2) sites, thus the effects of planar impurity can be studied, (ii) the hole concentration is believed to remain almost unaltered when divalent Cu(2) is substituted by divalent Zn, enabling one to look at the effects of disorder at almost the same hole content [5,6], and (iii) Zn reduces $T_c$ most effectively and therefore, provides a means to look at the low temperature behavior of various "normal state" properties as superconductivity is suppressed [7,8].

---

[*] Tel: +88-(0)721-750041/4102; Fax: +88-(0)721-750064; E-mail: salehnaqib@yahoo.com

In this paper we have reported the effect of Zn alloying on the hole content, p, and the oxygen deficiency, $\delta$, for a series of high-quality epitaxial c-axis oriented thin films and sintered $Y_{1-x}Ca_xBa_2(Cu_{1-y}Zn_y)_3O_{7-\delta}$ compounds over a wide variety of composition (x, y, and $\delta$). Ca substitution enabled us to overdope Y123 significantly as $Ca^{2+}$ in place of $Y^{3+}$ added holes in the $CuO_2$ planes, independent of the state of oxygenation in the $CuO_{1-\delta}$ chains [9-11]. In this study we have annealed a series of $Y_{1-x}Ca_xBa_2(Cu_{1-y}Zn_y)_3O_{7-\delta}$ under various annealing conditions and measured the changes in mass (giving the changes in the oxygen deficiency), room-temperature thermopower, *S[290K]*, and resistivity, $\rho(T)$. From the analysis of the room-temperature thermopower data, the temperature derivative of resistivity, $d\rho/dT$ at 300K, and the thermo-gravimetric analysis (TGA), we have found that annealing under identical conditions makes the Zn-substituted compounds more oxygen deficient compared to the ones without Zn.

**2. Experimental samples and results**

Polycrystalline single-phase sintered $Y_{1-x}Ca_xBa_2(Cu_{1-y}Zn_y)_3O_{7-\delta}$ were synthesized by standard solid-state reaction method using high-purity ( > 99.99%) powders. The details of sample preparation and characterization (using X-ray diffraction, EPMA, ac susceptibility, and resistivity measurements) can be found in ref. 9. High-quality c-axis oriented thin films of $Y_{1-x}Ca_xBa_2(Cu_{1-y}Zn_y)_3O_{7-\delta}$ were fabricated on $SrTiO_3$ substrates using pulsed laser deposition (PLD). Details of PLD, characterization, and oxygenation of the films can be found in refs. 9 and 12. Oxygen deficiencies were varied by quenching the samples from higher temperatures under different oxygen partial pressures into liquid nitrogen (LN$_2$). Samples of mass ~ 1gm were used to detect the changes in mass after each quenching. As we have substituted two different types of atoms, Ca and Zn in Y123, and we are interested in the possible role played by Zn on oxygen deficiency and hole content, samples with fixed amount of Ca were annealed together, having different amount of Zn. The annealing sequences employed for all the samples are given in Table-1. For the thin film compounds, it was not possible to detect the changes in δ-values from the weight changes, instead we have taken it to be identical to the ones for the sintered samples (used before as targets for the respective films) with same composition. Hole content for each sample was determined from the value of *S[290K]*. The room-temperature thermopower is a very convenient and accurate measure of the number of added hole carriers per CuO$_2$ plane, p, in cuprates [13] even in the presence of imperfections and strong in-plane scatterers like Zn [2,6,7,13,14]. *S[290K]* is also independent of the crystalline state of the sample [9-13]. p versus δ behaviour and *S[290K]* were found to be identical for crystalline films and sintered compounds, providing support to the assumption that for a given composition of the sample and annealing prescription, oxygen deficiency is the same for sintered and single crystalline compounds. We have obtained similar weight changes for Zn-free and various Zn-substituted samples (irrespective of Ca content). Therefore, as far as the *changes in δ-values* are concerned, identical annealing yields the same result. The above annealing treatments were done following earlier works by Tallon [15], Loram *et al.* [16], Cooper *et al.* [17], and De Silva [18]. Weight changes, and hence the oxygen deficiencies, found in this study agree very well with the previous studies [15,16]. It is important to notice in Table-I, that a higher annealing temperature is needed for the most Ca deficient sample (5%Ca-Y123) to get the same oxygen deficiency (δ) as that of 10%Ca-Y123. This annealing temperature for 5%Ca-Y123 has to be even higher for identical δ-values of 20%Ca-Y123. Thus, annealing samples with different amount of Ca at the same temperature masks, to some extent, the effect of Ca on the carrier concentration (p) due to unequal oxygen deficiencies. The general rule is, higher the carrier concentration (higher for samples with larger amount of Ca, when δ is constant), oxygen loss takes place at a lower temperature. This, we believe, has not probably been recognized by some earlier works [19,20] where questions has been raised regarding the role played by Ca in raising the overall hole concentration in $Y_{1-x}Ca_xBa_2Cu_3O_{7-\delta}$. Results of *S[290K]* versus Zn concentration of $Y_{0.80}Ca_{0.20}Ba_2(Cu_{1-y}Zn_y)_3O_{7-\delta}$ sintered compounds are shown in Fig. 1a for the different annealing conditions as described in Table - 1. Similar plots for 10% and 5%Ca-substituted samples are shown in Figs. 1b and 1c, respectively. It can be seen clearly from Figs. 1 that Zn does not change the thermopower significantly but there is a small but systematic change in *S[290K]* with Zn, namely *S[290K]* increases by a small amount in the Zn-substituted samples. This increment takes place with a small amount of Zn substitution (y = 1.5% to 3%) but remains almost unchanged (within the experimental error bars) for further Zn doping. Also this effect of Zn substitution is completely independent of the Ca content or the crystalline state of the samples. *S[290K]* has been widely used as an accurate measure of the number of added hole carriers per $CuO_2$ plane, therefore, these systematic changes in *S[290K]* would imply a small but non-monotonic change of p with Zn. At this point, one needs to ascertain the idea that *S[290K]* is indeed an accurate measure of p for the Zn-substituted samples by some other independent means. We have done this in the present study by investigating the changes in the values of temperature derivative of resistivity, dρ/dT at 300K for Zn-free and Zn-substituted samples as hole contents are varied. We have found dρ/dT at 300K to vary linearly with p for the Zn-free compounds under consideration. For these $Y_{1-x}Ca_xBa_2Cu_3O_{7-\delta}$ samples p-values were obtained from the well known parabolic $T_c(p)$ relation [21] given by $T_c(p) = T_{cmax}[1- 82.6(p-0.16)^2]$ (*S[290K]* yields identical values of p). Since, Zn merely adds a constant (T-independent) term to the resistivity [8,9,12], one would expect dρ/dT at 300K versus p behaviour to be identical for Zn-substituted and Zn-free compounds. This is what we have indeed found in this study. For the Zn-substituted samples p-values were solely determined from *S[290K]*. We have shown plots of dρ/dT at 300K versus p for a number of samples in Fig. 2. It is evident from Fig. 2 that *p-values for the Zn-substituted compounds are as reliable as those obtained from the widely used $T_c(p)$ relation* [21]. The linear trend shown in Fig. 2 is obtained by multiplying dρ/dT by the factors 3.5 and 1.3 for the 2%Zn-substituted 5%Ca-Y123 crystalline thin film and sintered 20%Ca-Y123 (with 1.5%Zn and Zn-free) samples respectively. These multiplicative factors have simple physical meanings, they represent mainly the degree of percolation for the resistive paths inside the samples [22-24], and agree quite well with previous studies [22-24]. Next, we have plotted the changes in the carrier concentration due to the different annealings for the $Y_{0.80}Ca_{0.20}Ba_2(Cu_{1-y}Zn_y)_3O_{7-\delta}$ compounds in Fig. 3. It is important to mention that, even though Zn-substituted samples have somewhat larger *S[290K]*, therefore, lower values of p, compared to the Zn-free ones for a given annealing, the changes in the values of *S[290K]* (and thus the changes in p) due to different annealings are almost identical, independent of Zn content. This is consistent with the fact that weight changes due to different oxygen annealings were also identical, irrespective of Zn concentration, as we have mentioned earlier. What might be the reason for this small difference in p between the Zn-free and the Zn-substituted samples after identical annealings? The possible reasons, we can think of, for this effect could be due to (i) strong ionic potential of $Zn^{2+}$, inducing changes in the charge transfer mechanism from the $CuO_{1-\delta}$ chains to $Cu(Zn)O_2$ planes and/or affecting the ab-plane charge dynamics in some other way (e.g., hole localization in the $Cu(Zn)O_2$ planes, reducing the concentration of mobile carriers) [25,26] or (ii) due to a change in the oxygen content induced by Zn. The first possibility seems unlikely considering a vast number of experimental results [4-6,8] indicating that Zn perturbs the electronic structure only locally and *S[290K]* is remarkably insensitive to local disorder [2,7,9,13]. We have investigated the second possibility in details in the present study. A plot giving *S[290K]* versus δ and the changes in p (obtained from *S[290K]*) due to the weight-losses from the annealings described in Table - 1 for a number of $Y_{1-x}Ca_xBa_2(Cu_{1-y}Zn_y)_3O_{7-\delta}$ compounds is shown in Fig. 4. Nearly parallel shifts in the values of *S[290K]* for samples with different amount of Ca (see Fig. 4) can be readily accounted for by taking into consideration the effect of Ca. Substitution of $Ca^{2+}$ for $Y^{3+}$ adds ~ x/2 holes per $CuO_2$ plane, irrespective of the value of δ [9-11]. There is also a parallel shift in the values of *S[290K]* for 20%Ca samples with 0%Zn and 3%Zn. For these compounds, *data shown in Fig. 4 have been presented with the assumption that the oxygen deficiency is independent of Zn-content, i.e., annealing under the same condition yields the same value for δ*. If this is true, then Fig. 4 would imply that Zn does change (decrease) the hole concentration, p, even though by a small amount. To check whether this assumption is true, one needs to anneal a pair of samples with same Ca content but with different Zn concentrations together and then totally deoxygenate those to detect the respective mass changes carefully. In this way one gets the absolute value of the oxygen deficiency after the annealing. We have performed this experiment with a number of pairs (one Zn-free and the other Zn-substituted) using thermo-gravimetric (TG) measurements. This is important, because as mentioned before, the changes in δ-values due to two different annealings yields the same value irrespective of Zn content, the uncertainty lies in the absolute values of δ at each annealing.

The *Stanton Redcroft STA 1500* thermal analyzer was used for TG measurements. This instrument showed a small drift in apparent mass with temperature. This drift was weakly dependent on the rate at which temperature was swept. Therefore, we have made background (empty-crucible) TG runs over the same temperature range, temperature sweep-rate, and gas flow environment as used in runs with samples before each experiment. This background was subtracted to correct the raw data. As a typical example, results of TGA for two compounds with compositions $Y_{0.80}Ca_{0.20}Ba_2Cu_3O_{7-\delta}$ and $Y_{0.80}Ca_{0.20}Ba_2(Cu_{0.97}Zn_{0.03})_3O_{7-\delta}$ after A1 are presented in this paper. We show the corrected weight-loss, Δm, (in %) versus temperature data for 0% and 3%Zn samples under flowing argon in Fig. 5. At 960°C the weight-loss reaches saturation and it is reasonable to assume that this is because of the complete loss of oxygen from the samples. For example the mass of $Y_{0.80}Ca_{0.20}Ba_2Cu_3O_{7-\delta}$ used for the TG run was 48.40 mg and after the TG run a weight-loss of 2.27% was recorded. This corresponds to an oxygen deficiency of 0.05 in the 48.40 mg sample as a result of the annealing A1. Similarly, for the sample $Y_{0.80}Ca_{0.20}Ba_2(Cu_{0.97}Zn_{0.03})_3O_{7-\delta}$ of mass 43.80 mg, the measured weight-loss was 2.08%, giving δ = 0.12, as a result of A1. Since both these samples have the same amount of Ca, this difference in δ for the two samples after identical annealing treatment should be due to Zn. Similar results were obtained for other pairs of samples with 10% and 5%Ca. Therefore, there is reason to believe that there is less oxygen in Zn-substituted samples, and based on our TG measurements the difference in the oxygen content between $Y_{0.80}Ca_{0.20}Ba_2Cu_3O_{7-\delta}$ and $Y_{0.80}Ca_{0.20}Ba_2(Cu_{0.97}Zn_{0.03})_3O_{7-\delta}$

samples is $\Delta\delta \sim 0.07$. We have re-plotted $S[290K]$ versus $\delta$ for $Y_{0.80}Ca_{0.20}Ba_2Cu_3O_{7-\delta}$ and $Y_{0.80}Ca_{0.20}Ba_2(Cu_{0.97}Zn_{0.03})_3O_{7-\delta}$ with and without the correction in $\delta$ found from TGA in Fig. 6. As can be seen from Fig. 6, the corrected data for 20%Ca-3%Zn-Y123 sample almost coincides (within the limits set by errors) with those for 20%Ca Zn-free compound. This indicates that the differences in $S[290K]$ between the Zn-free and Zn-substituted samples (see Figs. 4 and 6) are primarily because of unequal oxygen deficiencies.

## 3. Discussions and conclusions

From the analysis of TGA and $S[290K]$ it appears that Zn-substituted samples are more oxygen deficient than the Zn-free ones when annealed under identical temperatures and oxygen partial pressures. The possible physical reason for this effect is not clear to us. To get a definite answer one needs to address several puzzling questions. In Y123 holes are transferred from the $CuO_2$ plane through the apical oxygen from the $CuO_{1-\delta}$ chains. It is important to know how does an in-plane $Zn^{2+}$ (as $Zn^{2+}$ mainly substitutes the $Cu^{2+}$ in the $CuO_2$ plane) affects the binding energy of the oxygen in the $CuO_{1-\delta}$ chain. Also a difference in the oxygen levels of $\Delta\delta \sim 0.07$ was found for all the Zn-free and Zn-substituted samples irrespective of the amount of Zn substitution (in the range $y = 1.5\%$ to 6%). This is truly strange, it seems like a small amount of Zn makes the sample more oxygen deficient compared to the Zn-free one after identical oxygen annealing, but increasing Zn content further has little or no effect.

## Acknowledgements

The author would like to thank Dr. J. R. Cooper and Dr. J. W. Loram of University of Cambridge, UK, and Prof. J. L. Tallon of MacDiarmid Institute for Advanced Materials and Nanotechnology and Victoria University at Wellington, New Zealand, for their helpful comments and suggestions. The author would also like to thank the IRC in Superconductivity, University of Cambridge, UK, for funding this research.

**Figure captions** (Paper title: The effect of Zn substitution on the state of oxygen deficiency and hole concentration in $Y_{1-x}Ca_xBa_2(Cu_{1-y}Zn_y)_3O_{7-\delta}$ by S. H. Naqib)

Fig. 1: *S[290K]* versus Zn concentration data after different annealings for (a) sintered $Y_{0.80}Ca_{0.20}Ba_2(Cu_{1-y}Zn_y)_3O_{7-\delta}$, (b) sintered $Y_{0.90}Ca_{0.10}Ba_2(Cu_{1-y}Zn_y)_3O_{7-\delta}$, and (c) sintered and thin films of $Y_{0.95}Ca_{0.05}Ba_2(Cu_{1-y}Zn_y)_3O_{7-\delta}$.

Fig. 2: Temperature derivative of the resistivity, $d\rho/dT$, at 300K versus p for a number of sintered and thin film $Y_{1-x}Ca_xBa_2(Cu_{1-y}Zn_y)_3O_{7-\delta}$ samples. The dashed straight line was drawn as a guide to the eye.

Fig. 3: p versus Zn concentration for sintered $Y_{0.80}Ca_{0.20}Ba_2(Cu_{1-y}Zn_y)_3O_{7-\delta}$.

Fig. 4: Main panel: *S[290K]* versus oxygen deficiency, $\delta$, for a number of sintered $Y_{1-x}Ca_xBa_2(Cu_{1-y}Zn_y)_3O_{7-\delta}$ samples. Inset: p versus $\delta$ for the same compounds.

Fig. 5: Weight loss, $\Delta m$ (in %) versus temperature of $Y_{0.80}Ca_{0.20}Ba_2Cu_3O_{7-\delta}$ and $Y_{0.80}Ca_{0.20}Ba_2(Cu_{0.97}Zn_{0.03})_3O_{7-\delta}$ during the TGA (in argon) after the annealing A1.

Fig. 6: $S[290K]$ versus $\delta$ for $Y_{0.80}Ca_{0.20}Ba_2Cu_3O_{7-\delta}$ and $Y_{0.80}Ca_{0.20}Ba_2(Cu_{0.97}Zn_{0.03})_3O_{7-\delta}$. In the corrected data (circles), $\delta$ values for $Y_{0.80}Ca_{0.20}Ba_2(Cu_{0.97}Zn_{0.03})_3O_{7-\delta}$ are increased by 0.07.

**Table** (Paper title: The effect of Zn substitution on the state of oxygen deficiency and hole concentration in $Y_{1-x}Ca_xBa_2(Cu_{1-y}Zn_y)_3O_{7-\delta}$ by S. H. Naqib)

**Table 1**: Annealing and oxygen stoichiometry of $Y_{1-x}Ca_xBa_2(Cu_{1-y}Zn_y)_3O_{7-\delta}$.

| Annealing identity | Description | Oxygen stoichiometry ($\delta$) [± 0.02]* | |
|---|---|---|---|
| *(Sample composition: $Y_{0.80}Ca_{0.20}Ba_2(Cu_{1-y}Zn_y)_3O_{7-\delta}$; sintered)* | | | |
| AP20 | As prepared** | 0.010 [15] | |
| A1 | 300°C in $O_2$ for 96 hours | 0.057 | |
| A2 | 450°C in $O_2$ for 72 hours | 0.123 | |
| A3 | 520°C in $O_2$ for 48 hours | 0.196 | |
| A4 | 543°C in $O_2$ for 36 hours | 0.235 | |
| A5 | 565°C in $O_2$ for 36 hours | 0.283 | |
| A6 | 590°C in $O_2$ for 24 hours | 0.321 | |
| A7 | 615°C in $O_2$ for 24 hours | 0.366 | |
| A8 | 640°C in $O_2$ for 18 hours | 0.411 | |
| A9 | 670°C in $O_2$ for 18 hours | 0.444 | |
| A10 | 690°C in $O_2$ for 12 hours | 0.465 | |
| A11 | 550°C in 1%$O_2$ + 99%$N_2$ for 24 hours | 0.505 | |
| *(Sample composition: $Y_{0.90}Ca_{0.10}Ba_2(Cu_{1-y}Zn_y)_3O_{7-\delta}$ and $Y_{0.95}Ca_{0.05}Ba_2(Cu_{1-y}Zn_y)_3O_{7-\delta}$; sintered)* | | | |
| | | **10%Ca** | **5%Ca** |
| AP10/5 | As prepared (10%Ca)*; (5%Ca)*** | 0.010 [15] | 0.020**** |
| B1 | 350°C in $O_2$ for 72 hours | 0.030 | 0.030 |
| B2 | 400°C in $O_2$ for 72 hours | 0.057 | 0.048 |
| B3 | 450°C in $O_2$ for 48 hours | 0.085 | 0.080 |
| B4 | 500°C in $O_2$ for 36 hours | 0.132 | 0.123 |
| B5 | 550°C in $O_2$ for 36 hours | 0.188 | 0.173 |
| B6 | 600°C in $O_2$ for 24 hours | 0.272 | 0.258 |
| B7 | 650°C in $O_2$ for 24 hours | 0.379 | 0.340 |

*(Sample composition: c-axis oriented $Y_{0.95}Ca_{0.05}Ba_2(Cu_{1-y}Zn_y)_3O_{7-\delta}$; thin film)*

| | | |
|---|---|---|
| AP | As prepared [9,12] | 0.050***** |
| C1 | 500°C in $O_2$ for 2 hours | 0.122 |
| C2 | 550°C in $O_2$ for 2 hours | 0.171 |

* These δ-values are for Zn-free compounds

**All AP 20% and 10%Ca substituted samples were annealed *in-situ* at high oxygen partial pressure at 300°C

*** AP 5%Ca substituted samples underwent *in-situ* oxygen annealing at 700°C (8 hours) followed by annealing at 300°C (48 hours) and at 175°C (24 hours). Here the high-temperature (700°C) annealing was for the uniform oxygenation inside the bulk and the low-temperature (175°C) one was done to oxygenate the grain boundaries.

**** From thermo-gravimetric measurement

***** Estimated from the values of *S[290K]*

Fig. 1:

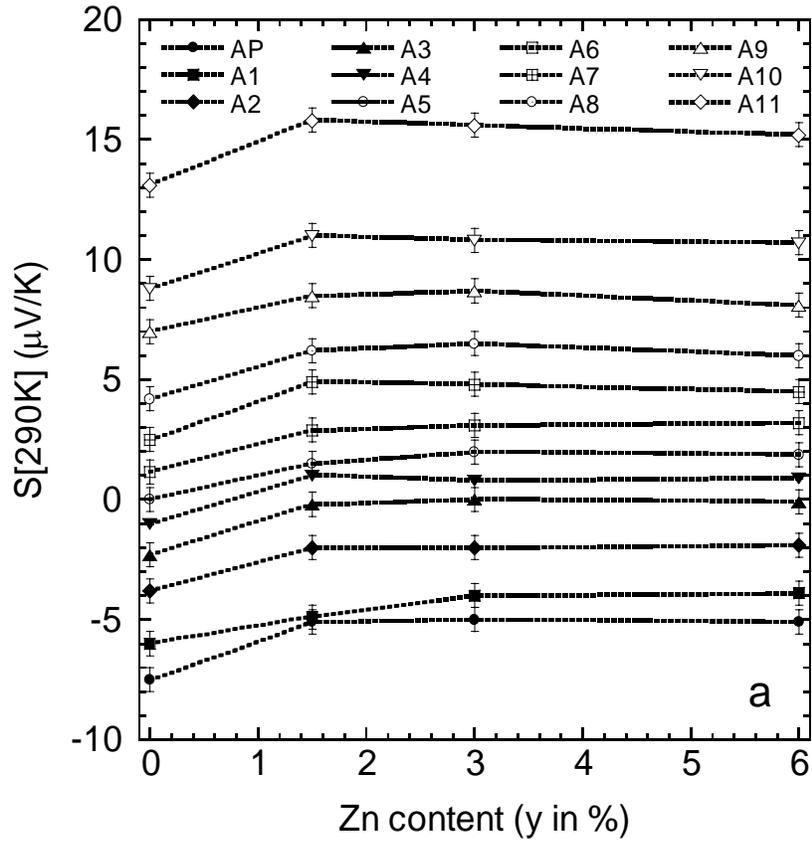

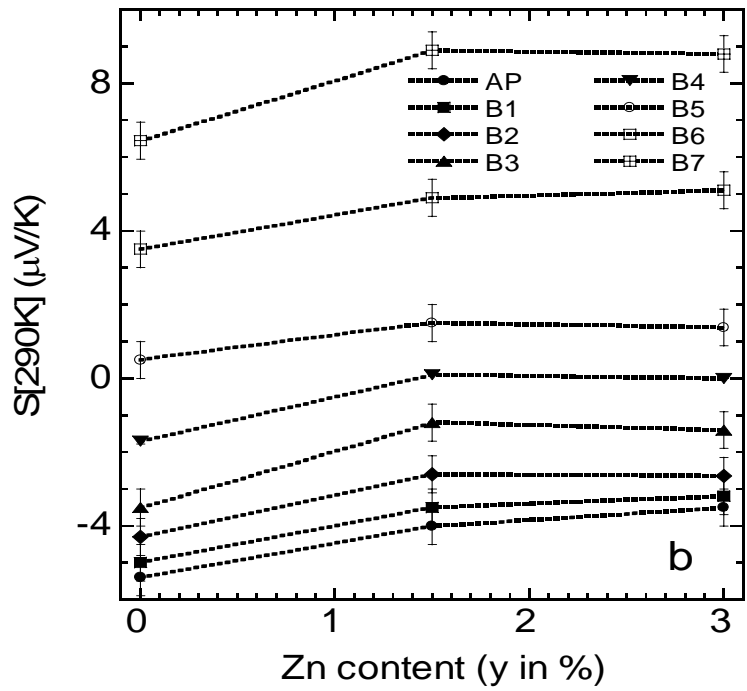

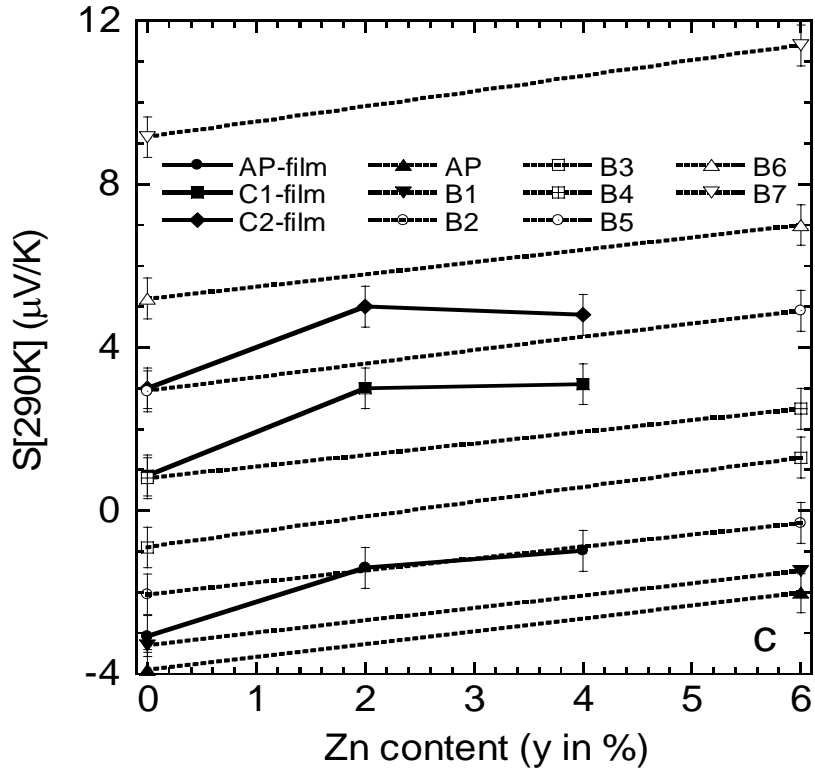

Fig. 2:

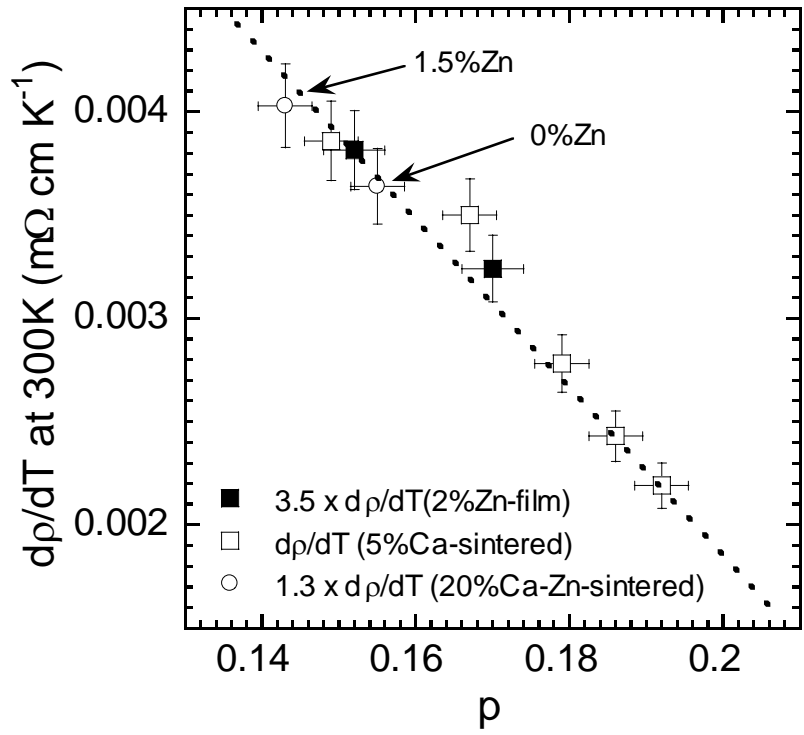

Fig. 3:

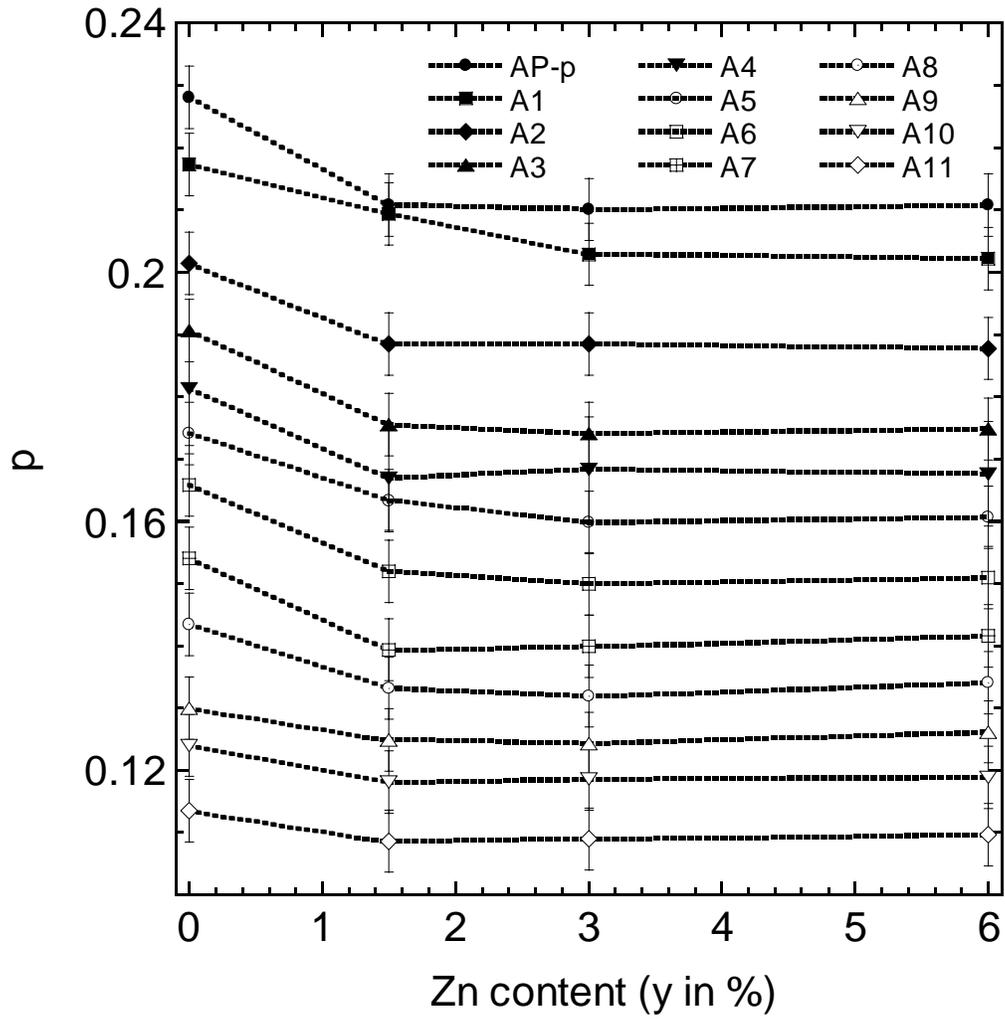

Fig. 4:

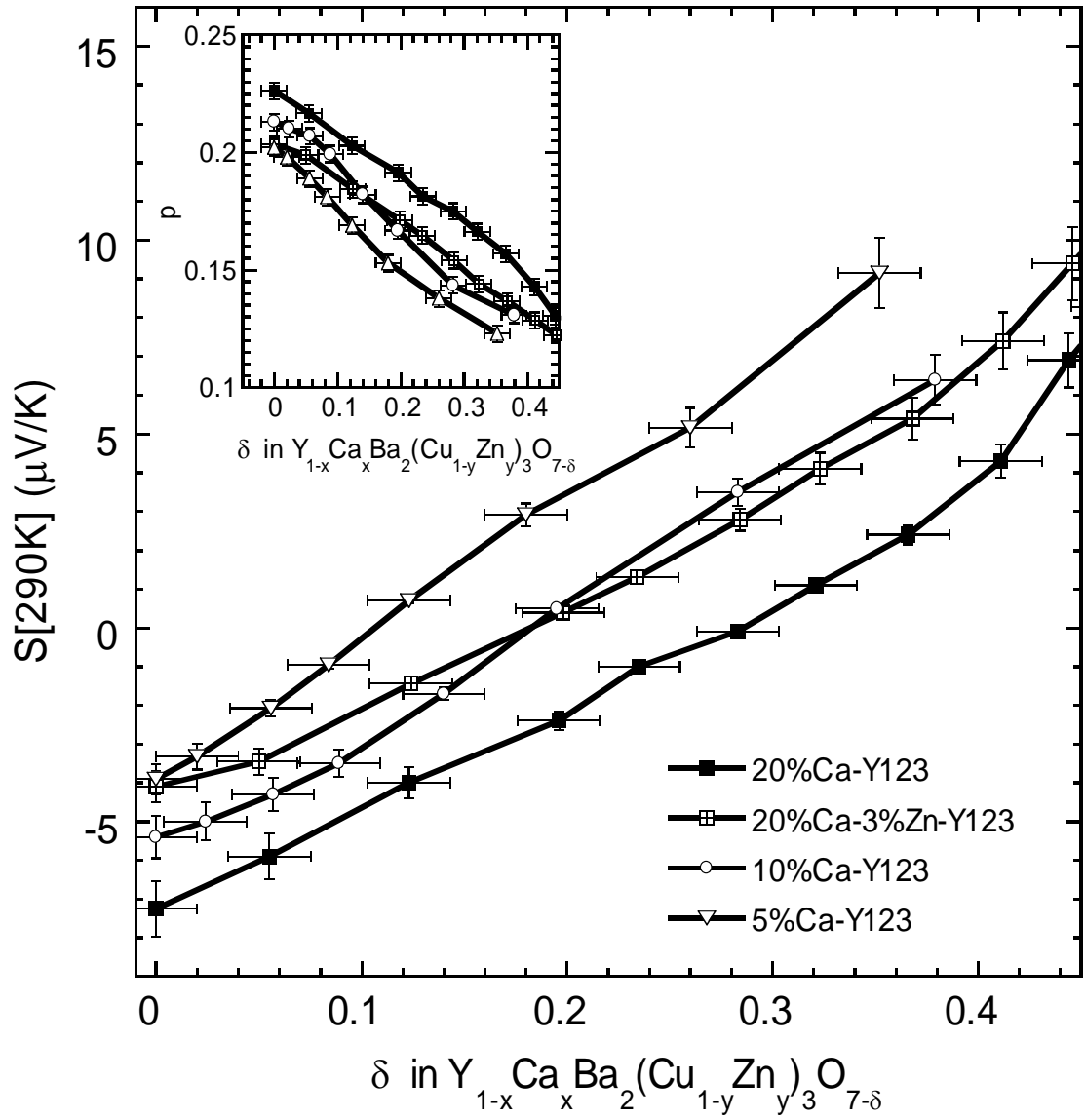

Fig. 5:

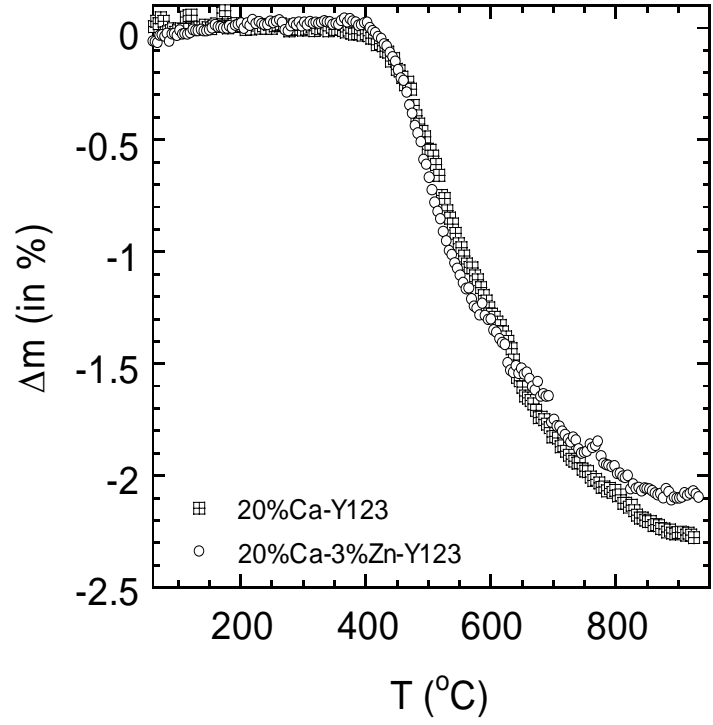

Fig. 6:

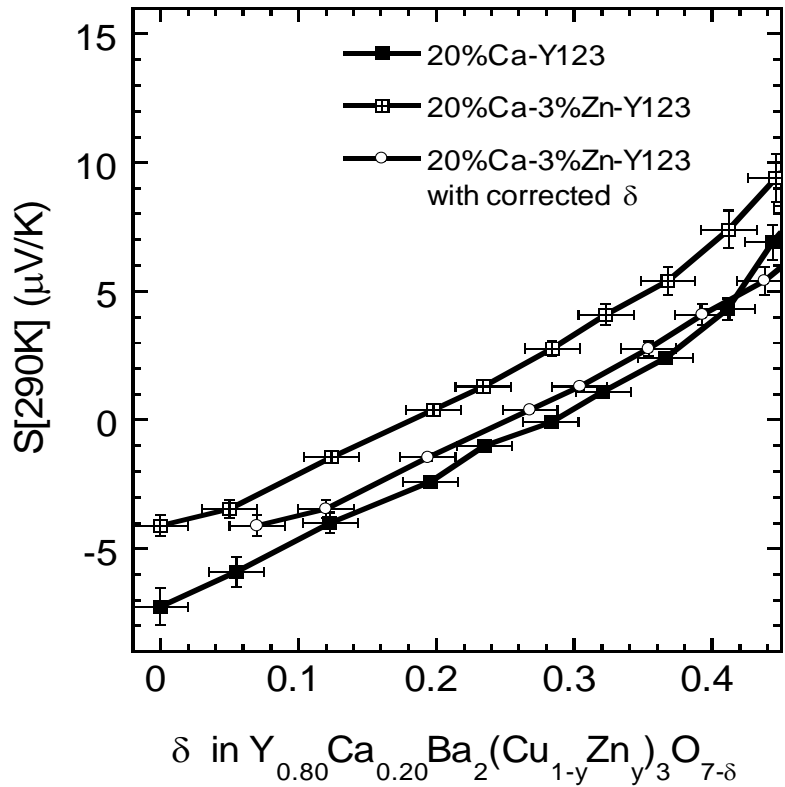